# Una paradoja hidrostática

**E. N. Miranda**
**CCT Mendoza - CONICET**
**5500 – Mendoza, Argentina**
**E-mail: emiranda@mendoza-conicet.gov.ar**

Se muestra que en un sistema de vasos comunicantes que cambia de un estado de equilibrio a otro, no se pueden satisfacer simultáneamente el principio de Pascal, el de conservación de la masa y el de la energía.

Palabras claves: hidrostática, conservación de la energía

It is shown that for a liquid in a connected vessel system, it is not possible to fulfill simultaneously Pascal´s principle, mass conservation and energy conservation when the system goes from an equilibrium state to another one.

Hay temas que se enseñan una y otra vez en los cursos de Física General y que aún tienen aspectos por descubrir. Es el caso de la paradoja hidrostática que se presenta en esta nota. Veremos que la aplicación simultánea de propiedades de vasos comunicantes, la conservación de la masa y de energía lleva a un resultado inesperado.

La situación experimental que se analiza es la descripta por la Figura 1. Se tiene una vasija provista de vasos comunicantes y llena con un fluido de densidad $\delta$. Para fijar ideas supondremos que tenemos tres vasos comunicantes y con áreas respectivas $A_1, A_2, A_3$. El fluido tiene inicialmente la misma altura $h_0$ en los tres vasos. A los efectos de llevar a cabo nuestro experimento supondremos que hay una placa de peso despreciable sobre el fluido en cada uno de los vasos.

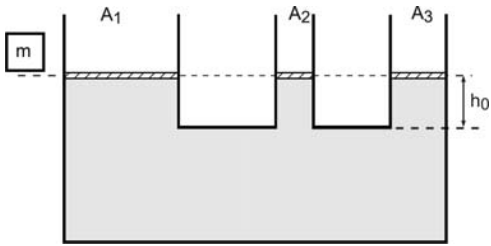

*Figura 1: Inicialmente el sistema se encuentra en la situación descripta: el nivel de líquido en los vasos comunicantes es $h_0$ y hay una masa m externa que está a la misma altura.*

Adicionalmente al fluido, inicialmente hay una masa $m$ a una altura $h_0$, fuera de la vasija.

La situación final es la mostrada en la Figura 2. La masa $m$ ha sido colocado sobre el fluido en el vaso 1 por lo que su altura ha disminuido y ahora es $h_1$. Para que se verifique la igualdad e presiones en los puntos $a$, $b$ y $c$ que están a igual nivel, el líquido sube en los vasos 2 y 3 hasta alcanzar la altura $h_2$ y $h_3$ que en principio podrían diferir.

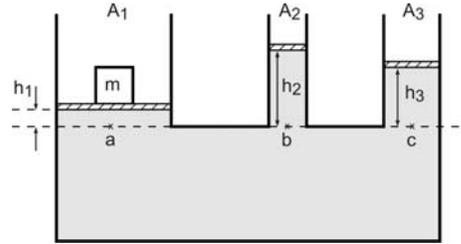

*Figura 2: Este es el estado final del sistema una vez que se ha colocado la masa m en el primer vaso comunicante. Las alturas de los líquidos en los tres vasos son ahora $h_1$, $h_2$ y $h_3$.*

Nuestro objetivo es estudiar este sistema teniendo en cuenta que hay tres condiciones que deben cumplirse simultáneamente.

En primer lugar, las presiones en los puntos $a$, $b$ y $c$ deben ser iguales (principio de Pascal); en términos matemáticos tiene que ser $p_a = p_b = p_c$. Entonces se puede escribir:

$$\frac{mg}{A_1} + \delta g h_1 = \delta g h_2 \qquad (1a)$$

$$\delta g h_2 = \delta g h_3. \qquad (1b)$$

En segundo lugar, la masa se debe conservar, por lo tanto, la cantidad de fluido que

sale del vaso 1 debe ser igual al incremento en las masas de fluido de los vasos 2 y 3. Entonces:

$$A_1(h_0 - h_1) = A_2(h_2 - h_0) + A_3(h_3 - h_0). \quad (2)$$

Hay que hacer notar que tenemos un sistema lineal con tres incógnitas, $h_1$, $h_2$ y $h_3$ y tres ecuaciones: la (1a), (1b) y la (2). Entonces el problema tiene solución única y se pueden despejar las nuevas alturas de la columna de líquido. Se encuentra así que:

$$h_1 = h_0 - \frac{A_2 + A_3}{A_1(A_1 + A_2 + A_3)} \frac{m}{\delta}$$
$$h_2 = h_0 + \frac{1}{(A_1 + A_2 + A_3)} \frac{m}{\delta} \quad (3)$$
$$h_3 = h_0 + \frac{1}{(A_1 + A_2 + A_3)} \frac{m}{\delta}$$

Finalmente, uno podría conjeturar que debería conservarse la energía. Hay que recordar que la energía potencial de una columna de líquido de altura $h$ y sección $A$ es $\frac{1}{2} \delta A h^2$. En consecuencia, tenemos las energía inicial $E_i$ y final $E_f$ dadas por:

$$E_i = mgh_0 + \tfrac{1}{2}\delta g A_1 h_0^2 + \tfrac{1}{2}\delta g A_2 h_0^2 + \tfrac{1}{2}\delta g A_3 h_0^2 \quad (4a)$$

y

$$E_f = mgh_1 + \tfrac{1}{2}\delta g A_1 h_1^2 + \tfrac{1}{2}\delta g A_2 h_2^2 + \tfrac{1}{2}\delta g A_3 h_3^2 \quad (4b)$$

La sorpresa surge a partir de las ecuaciones (4a) y (4b). En nuestro análisis no se ha introducido la viscosidad del fluido, por lo tanto cabría esperar que se conserva la energía; sin embargo no es así. Si calculamos la diferencia entre las energías inicial y final tenemos:

$$\Delta E = E_f - E_i$$
$$= mg(h_1 - h_0) + \tfrac{1}{2}\delta g A_1(h_1^2 - h_0^2) + \quad (5)$$
$$+ \tfrac{1}{2}\delta g A_2(h_2^2 - h_0^2) + \tfrac{1}{2}\delta g A_3(h_2^2 - h_0^2)$$

Si se reemplazan los valores encontrados para $h_1$, $h_2$ y $h_3$ en (3) resulta:

$$\Delta E = -\frac{1}{2}\frac{m^2 g}{\delta}\frac{A_2 A_3}{A_1(A_1 + A_2 + A_3)} \quad (6)$$

Es claro que $\Delta E$ es siempre negativo. ¿Qué pasó con el principio de conservación de la energía? Dónde está la energía faltante?

La respuesta a esta paradoja está en que tácitamente hemos supuesto que el líquido es viscoso. Al colocar la masa sobre el líquido del vaso 1, el sistema queda fuera de equilibrio y comienza a oscilar. Para que se restablezca nuevamente el equilibrio y se alcance la condición final mostrada en la Figura 2, es necesario que las oscilaciones se amortigüen. Y para que tal cosa suceda, se tiene que disipar energía debido a la viscosidad del líquido. Esta situación es análoga a la conocida paradoja de los dos capacitores cargados que se conectan a través de un conductor presuntamente sin resistencia [1, 2] y al hacerlo se pierde energía.

En síntesis, al considera el sistema mostrado en las Figuras 1 y 2, ingenuamente se supondría que debe cumplirse simultáneamente tres condiciones: igualdad de presión a igual profundidad, conservación de la masa y conservación de la energía. Sin embargo no es así. Para poder pasar de la situación inicial a la final, tiene que haber disipación viscosa en el líquido y por lo tanto la energía no se conserva. Queda como ejercicio para el lector ver que para otras configuraciones más complejas de vasos comunicantes la situación es la misma.

**Referencias:**